\documentstyle[aps,prb,epsf,graphicx]{revtex}
\begin{document}

\newcommand{\lessim}{\mbox{\tiny$\mbox{\normalsize$<$}\atop
\mbox{\normalsize$\sim$}$}}
\newcommand{\grtsim}{\mbox{\tiny$\mbox{\normalsize$>$}\atop
\mbox{\normalsize$\sim$}$}}

\title{Superconducting proximity effect in clean ferromagnetic layers}
\author{M.~Zareyan$^{1,2}$, W.~Belzig$^{3}$, and Yu.~V.~Nazarov$^{2}$}
\address{
  $^1$ Institute for Advanced Studies in Basic Sciences, 45195-159,
  Zanjan, Iran\\
  $^2$ Department of Applied Physics and Delft Institute of
  Microelectronics and Submicrontechnology,\\ Delft University of
  Technology, Lorentzweg 1, 2628 CJ Delft, The Netherlands\\
  $^3$ Department of Physics and Astronomy, University of Basel,
  Klingelbergstr. 82, 8056 Basel, Switzerland}
\date{\today}
\maketitle

\begin{abstract}
  We investigate superconducting proximity effect in clean
  ferromagnetic layers with rough boundaries. The subgap density of
  states is formed by Andreev bound states at energies which depend on
  trajectory length and the ferromagnetic exchange field. At energies
  above the gap, the spectrum is governed by resonant
  scattering states. The resulting density of states, measurable by
  tunneling spectroscopy, exhibits a rich structure, which allows to
  connect the theoretical parameters from experiments.
\end{abstract}

\pacs{PACS numbers: 74.50.+r, 74.80.-g, 75.70.-i}

\section{Introduction}
Investigating of superconducting proximity effect in normal
systems has a long history back to the experiments of McMillan and
Rowell\cite{rowell}. Their tunneling spectroscopy measurements in
normal metals connected to a superconductor revealed strong
modifications of the density of states (DOS) caused by the induced
superconducting correlations. These results were understood in the
tunneling model of McMillan\cite{mcmillan:68}. He first noted that
the changes in the DOS of the normal metal occur on a scale
$E_{\text{Th}}$, which in his model is identified with inverse
escape time of a quasi-particle in the normal metal. Naively, one
would have expected $\Delta$, the superconducting gap parameter,
to play the dominant role, which turned out to be not the case.
Furthermore, if $E_{\text{Th}}\ll \Delta$, the energy gap in the
superconductor plays no role at all and the DOS in the normal
metal has a gap of the order of $E_{\text{Th}}$. We can understand
these observations by noting that in clean normal metal films the
electronic properties are determined by so called Andreev bound
states \cite{andreev}. These are bound electron-hole pairs living
on trajectories, which start and end at the superconductor. The
characteristic energy scale in this case is $v_{\text{F}}/d$,
again the inverse escape time. In recent years experiments became
possible, in which the density of states was resolved {\em
locally} on a sub-$\mu$m scale. For example, the dependence of the
tunneling DOS on the distance from the superconductor in normal
metals has been measured by Gu\'eron {\it et al.}\cite{gueron}
using additional tunnel junctions. These results have been
successfully explained within the quasi-classical theory in the
diffusive limit.\cite{belzig:96} Other experiments made use of low
temperature scanning tunneling microscopes to resolve spatially
the DOS of small droplets of normal metal on the surface of a
superconductor.\cite{tessmer} Nowadays these types of experiments
are becoming a standard technique.\cite{millo,courtois,vinet}

The question of the proximity effect in the presence of a spin
splitting is currently heavily investigated. In particular the
influence of a superconductor on transport properties of a
ferromagnet is under debate\cite{falko,jedema,belzig:00},in which
case the proximity effect is negligible. It is, however, natural
to address the question of the influence of an exchange field on
the proximity density of states. In fact, this question was
already addressed experimentally a while ago by Gallagher {\it et
al. } \cite{tedrow}.  They observed a spin splitting of the DOS in
thin normal layers in a parallel magnetic field. New experimental
developments,\cite{aprili:01} exploring the proximity effect on a
nanometer scale demand new theoretical models, beyond the simple
tunneling model of McMillan. In the present paper we investigate a
new model, suitable for these experiments.

The motivation stems not only from fundamental question of the
coexistence of ferromagnetism and superconductivity, but also because
interesting applications of ferromagnet-superconductor (FS) hybrid
structures have been proposed. We only mention here the potential use of
SFS-contacts in the construction of quantum computers. SFS-junctions
are candidates for all-electronic $\pi$-junction, which are needed in some
proposals for solid-state qubits. \cite{blatter,mooij}

We also note, that many surprising effects in normal metal-superconductor
heterostructures have been found, both experimentally and theoretically.
The question, how these effects are modified by the presence of
ferromagnetism, is of high interest. For example, making use of
mechanically controlled break junctions, the conduction channel content
of single atoms has been determined \cite{scheer:97}. These were shown
to depend on the chemistry of the atom only\cite{scheer:98}. It was
however crucial to have superconducting contacts in these experiments.
Therefore, so far only superconducting elements or normal metals, with
the help of the proximity effect, have been
investigated.\cite{scheer:01} It is reasonable to expect that
experiments with ferromagnetic materials will be performed in the near
future.

Long time ago Larkin and Ovchinnikov \cite{LO} and Fulde and
Ferrel \cite{FF} independently predicted, that in the presence of
an exchange field $h$ (for instance in magnetic superconductors) a
specific superconducting state can be formed, in which Cooper
pairs have a non-zero momentum due to spin splitting. The momentum
of Cooper pairs is given by $2h/v_{\text F}$, which, in the
singlet state, results from the difference between up and down
spin band Fermi momenta. The resulting LOFF state is qualitatively
different from the homogeneous zero-momentum state. Due to a
varying superconducting phase, the order parameter has an
oscillatory spatial modulation containing nodes, in which the
phase changes by $\pi$. The LOFF state has never been observed in
bulk superconductor, but there are recently evidences for
detecting an induced LOFF state in heterostructures of
ferromagnets and superconductors. Many works have investigated the
thermodynamics properties of FS-multilayers. Radovic {\it et
al.}\cite{Radovic} have predicted oscillation of the
superconducting critical temperature $T_c$ as a function of the
thickness of the attached F-layer. The experimental evidence for
these $T_c$ oscillations is not, however, conclusive.\cite{strunk}
The reason for this may, for example, result from a bad quality of
the FS-interface.\cite{aarts}

The most recent experiments have concentrated on other properties
of FS-layers. Ryasanov {\it et al.} \cite{ryasanov} measured the
temperature dependence of the critical current in SFS Josephson
junctions with thin F-layer and have found a non-monotonic
temperature dependence. This behavior can be understood in terms
of a $\pi$ phase shift due to the exchange field, which occurs for
certain values of the thickness of the F-layer, as first was
predicted by Bulaevskii {\em et  al.}\cite{bulaevski}.  An
indirect proof of the $\pi$ phase shift has been made by Kontos
{\it et al.},\cite{aprili:01}, who studied the density of states
in thin ferromagnetic films contacted by a superconductor.  They
observed an oscillatory behavior of the induced superconducting
correlation for layers of different thickness, which was
attributed to influence of the exchange field. In Ref.
\cite{ZBN:01}, we have shown that these experimental findings
could be explained by a model of a ballistic ferromagnetic layer
with rough boundaries. The best agreement was obtained in the
limit of large $h/\Delta$ and small interface transparency $T$.

In the present paper, we study the proximity DOS in a clean
ferromagnetic layer on top of a superconductor in the full
parameter range. Within the ballistic quasi-classical formalism we
obtain, that the DOS for energies below the superconducting gap
$\Delta$ is completely specified by the length distribution of the
classical trajectories inside the ferromagnet
(Sec.~\ref{sec:model}). The DOS for energies above the gap is also
expressed in terms of the length distribution of the trajectories.
The length distribution depends on the geometrical properties of
the attached ferromagnet and the connecting boundaries. In
Sec.~\ref{sec:distribution} we specify the classical length
distribution for our particular case of the F-film geometry
depicted in Fig.~\ref{zbnfig1}.  We assume, that the boundaries of
the F-film are disordered, leading to complete diffusive
reflection of the quasi-particles from these boundaries. We also
take into account band mismatch and disorder at the FS-interface,
which leads to an enhanced backscattering from this interface. For
simplicity, we assume a single value of the FS-interface
transparency $T$.  With the calculated distribution, the DOS at
all energies is obtained as a function of the superconducting gap
$\Delta$, the exchange field $h$, the thickness of the F-layer $d$
and the transparency $T$.

We analyze the DOS for different regimes of $h/\Delta$. It shows
the interplay between ferromagnetism and superconductivity
depending qualitatively on the thickness $d$. For example, a weak
exchange field leads to a spin-splitting of the DOS, which results
in a distinctive low energy peak in the total DOS. In addition
there is an overall suppression of the superconducting features of
the DOS with increasing $h$. At higher exchange fields, the DOS
shows as a signature of the exchange splitting an oscillatory
behavior as a function of the layer thickness.  This oscillation
of DOS was observed in the experiments \cite{aprili:01}. Our
findings are summarized in the following list:

\begin{itemize}
\item $h=0$ (Sec.~\ref{subsec:normal}): Andreev levels are governed by
  the distribution of trajectory lengths, which only depends on the
  geometric properties of the sample. At small energies $E\ll
  v_{\text{F}}/d$ the DOS is strongly suppressed, originating from the
  exponential suppression of long trajectories. The DOS at larger
  energies reflects the length distribution. In our model it display a
  multiple peak structure, resulting from multiple reflections at the
  SN-interface. The resulting minigap correspond to the gap found in
  a calculation including impurity scattering.\cite{pilgram:00}

\item $h/\Delta < 1$ (Sec.~\ref{subsec:weak}): A small exchange field
 'splits' the DOS for spin up and down quasiparticles, i.~e.~the
  total DOS is more or less a superposition of 'normal' DOS's at
  energies $E\pm h$. Accordingly, the former minigap in the DOS is
  destroyed. It only remains a dip in the DOS shifted to finite
  energies. The density of states at the Fermi level
  approaches the normal state values in an oscillatory way, i.~e.
  overshooting the normal DOS for certain values of $h$.

\item $h/\Delta \gtrsim 1$ (Sec.~\ref{subsec:weak2}): The superconducting
  features of the DOS are stronger suppressed. The former peaks at $\pm\Delta$
  are inverted into dips for thicker layers. Above the gap peaks at $E=\pm h$
  appear as the signature of resonant transmission through the ferromagnetic
  film.  For thin layers features at $\pm h$ are absent and the DOS approaches
  a BCS-form.

\item $h/\Delta \gg 1$ (Sec.~\ref{subsec:strong}): For layers with
  $d\grtsim v_{\text{F}}/h$ the DOS exhibits coherent oscillation, i.~e.
  the form of the DOS difference from the normal state value becomes
  independent of $d$. The amplitude and sign, however, depends on the
  thickness.  Only for very low thicknesses $d\ll Tv_{\text{F}}/h$
  the DOS approaches the BCS form.
\end{itemize}

In Sec.\ref{subsec:map} we condense our results into a map of the
proximity DOS. Finally we present some conclusions in
Sec.~\ref{sec:conclusion}.

\section{Model and basic equations}
\label{sec:model}

\begin{figure}[h]
  \begin{center}
    \includegraphics[width=6cm,clip=true]{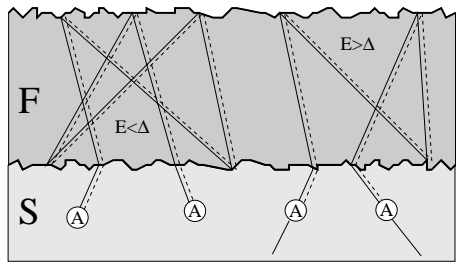}
    ~\\ ~\\
    \caption[]{Schematic drawing of our model of a ferromagnetic film (F) topping a
    superconductor (S).  Typical classical trajectories are also indicated.  We
    distinguish two processes. At energies $E$ below the superconducting
    gap $\Delta$, quasiparticles are confined to the film by Andreev reflection
    (indicated by the white circles). An examples for this process is given by
    the left trajectory. For $E>\Delta$ Andreev reflection is
    incomplete and the quasiparticles states in F are formed by scattering
    resonances, symbolized in the right process.}
   \label{zbnfig1}
 \end{center}
\end{figure}

The system we study is sketched in Fig.~\ref{zbnfig1}. A
ferromagnetic layer (F) of thickness $d$ is connected to a
superconducting bank (S) on one side and bound on the other side
by an insulator or vacuum. F is characterized by an exchange
splitting, which we take into account as mean field $h$ in the
Hamiltonian.  The thickness $d$ is larger than the Fermi wave
length $\lambda_{{\text{F}}}$ and smaller than the elastic mean
free path $\ell_{\text{imp}}$, which allows for a quasi-classical
description \cite{eilenberger} in the clean limit. We apply the
Eilenberger equation in the clean limit
\begin{equation}
  -i{\bf{v}_{\text{F}}}{\bf \nabla}
  \hat{g}_{\sigma}(\omega,{\bf{v}_{\text {F}}},{\bf r}) =
  \left[(E+\sigma h({\bf r}))\hat{\tau}_3-i\hat\tau_2\Delta({\bf r}) ,
  \hat{g}_{\sigma}(\omega,{\bf{v}_{\text {F}}},{\bf r})\right] .
  \label{Eil}
\end{equation}
The matrix Green's function for spin $\sigma$ has the form
\begin{equation}
  \hat{g}_{\sigma}=\left(
    \matrix{
      g_{\sigma} & f_{\sigma}\cr
      f^{\dag} _{\sigma} & -g_{\sigma}\cr}
  \right)\,.
\end{equation}
It depends on energy $E$, the direction of the Fermi velocity
${\bf{v}_{\text {F}}}$ and the coordinate ${\bf r}$. Here
$\hat{\tau}_i$ denote the Pauli matrices, $\Delta (\bf{r})$ is the
superconducting pair potential (taken as real), and $\sigma$
($=\pm 1$) labels the electron spin. The matrix Green's functions
obey the normalization condition $\hat{g}^2_\sigma=1$. Inside then
F-layer $h$ is constant and $\Delta=0$. We neglect a depression of
the pair potential close to the FS interface, thus
$\Delta({\bf{r}}) = \text{const.}$ inside the superconductor,
which applies in the case of a bad contact between the ferromagnet
and the superconductor. Strictly speaking, we would have to
include an elastic collision term in (\ref{Eil}), even in the
limit $\ell_{\text{imp}}\gg d$. However, changes in the spectrum
due to this term are limited to small energies $\lessim
v_{\text{F}} / \ell_{\text{imp}}\ll \text{min}(v_{\text{F}} /
d,h)$\cite{pilgram:00}, which are negligible in all cases we
study, except for the case $h=0$.  Disorder in the superconductor
can be neglected in the limit of small interface transmission,
which we mostly assume.

We have to solve Eqs.~(\ref{Eil}) along each classical trajectory
with length $l$ in F, that comes from the superconductor and ends
there.  As boundary conditions the solutions approach the bulk
values of $\hat{g}_{\sigma}$ at the beginning and the end of a
trajectory deep inside the superconductor.  These are given by
$\hat{g}_{\sigma}(\text{bulk}) = (-iE \hat{\tau_3} + \Delta
\hat{\tau_1})/\sqrt{\Delta^2-E^2}$.\cite{KO} It turns out that on
a trajectory inside F the normal Green function $g_{\sigma}$ is
constant.  It depends only on the length of that trajectory $l$
and is given by
\begin{eqnarray}
 g_{\sigma}=\tanh{[(-iE-i\sigma
 h)l/v_{\text{F}}+\arcsin{(-iE/\Delta)}]}.
 \label{g}
\end{eqnarray}
To find the density of states {\em per} trajectory, we have to
calculate
\begin{eqnarray}
   N(E,l)=\frac{N_0}{2}\sum\limits_{\sigma=\pm1}
  {\text{Re}}{g}_\sigma(E+i0,{\bf{v}_{\text{F}}},\bf{r})\,.
  \label{dosgeneral}
 \end{eqnarray}

As the result we obtain for energies below the gap ($|E|\leq \Delta$)
\begin{eqnarray}
    N(E,l)=\frac{N_0}{2}\sum\limits_{\sigma=\pm1}
  \frac{\pi v_{{\text{F}}}}{|E+\sigma h|}
  \sum_{n=-\infty} ^{\infty}\delta (l-l_n)\,,
  \label{trajdosb}
\end{eqnarray}
where
\begin{eqnarray}
  l_n=\frac{v_{{\text{F}}}}{E+\sigma h}
  \left(n\pi+\arccos(E/\Delta)\right)\,.
  \label{quantcond}
\end{eqnarray}
Above the gap ($|E|> \Delta$) we find
\begin{eqnarray}
   N(E,l)=\frac{N_0}{2}\sum\limits_{\sigma=\pm1}
   \sum_{n=-\infty} ^{\infty} \frac{{\text {acosh}}|E/\Delta |}
   {[(E+\sigma h)l/v_{\text{F}}-n\pi]^2
  +{({\text {acosh}}|E/\Delta|)}^2}\,.
 \label{trajdosa}
\end{eqnarray}
Here $N_0$ is the density of states at the Fermi level in the normal
state. Eqs.~(\ref{trajdosb}) express the fact that the density of states
below $\Delta$ is a sum of $\delta$-peaks resulting from Andreev bound
states of electrons of $E\geq 0$ (positive $n$'s) and holes of $E<0$
(negative $n$'s). The energies also follow from the quasi-classical
quantization condition $l=l_n$.

The total DOS is obtained by averaging the expressions (\ref{trajdosb})
and (\ref{trajdosa}) over all classical trajectories. Denoting the
trajectory length distribution by $p(l)$ and using (\ref{trajdosb}), we
find for the sub-gap DOS
\begin{eqnarray}
 \nonumber
  N(E)&=&\int dl p(l)N(E,l) \\
  \label{eq:dosb}
  &=&\frac{N_0}{2}\sum_{\sigma=\pm 1}
  \frac{\pi v_{{\text{F}}}}{|E+\sigma h|}
  \sum _{n=-\infty} ^{\infty} p(l_n) \ \ \ {\text{for}} \ \ |E|\leq \Delta.
\end{eqnarray}
This formula is a general result for the sub-gap density of states of a
quasi-ballistic metal connected to a superconductor. It is
completely specified by the length distribution of classical
trajectories, which depends only on the geometrical properties of the
attached ferromagnet and the surrounding boundaries.

Averaging expression (\ref{trajdosa}) over $p(l)$ the total DOS
for energies above the gap is
\begin{eqnarray}
   N(E)=\frac{N_0}{2}\sum\limits_{\sigma=\pm1}
   \sum_{n=-\infty} ^{\infty} \int dl p(l)\frac{{\text {acosh}}|E/\Delta |}
   {[(E+\sigma h)l/v_{\text{F}}-n\pi]^2
  +{({\text {acosh}}|E/\Delta|)}^2}
  \ \ \ {\text{for}} \ \ |E|> \Delta,
\label{eq:dosa}
\end{eqnarray}
The absence of discrete bound states reflects the fact, that the Andreev
reflection at energies above the gap is incomplete. Therefore, the quasi
particle states in the ferromagnet are determined by 'scattering
resonances' of quasiparticles incident from the superconductor.

\section{Distribution of the trajectory length}
\label{sec:distribution}

Now we specify the length distribution for our particular case. We
model the F-layer by a weakly disordered thin film bounded by a
rough surface to the insulator and a rough FS-interface of average
transparency $T$.  A typical classical trajectory is depicted in
Fig ~\ref{zbnfig1}. An electron coming from the bulk of S enters
into the F-layer and after several reflection from the insulator
and the FS-interface returns to the S-bank, where it is Andreev
reflected as a hole and transverses the trajectory in the opposite
direction. Thus, the building block of a trajectory is the segment
between two successive reflections from S. The number of blocks
which form the total trajectory depends on the transparency of the
interface, i.~e., it is roughly $\sim 1/T$.

As first step, we consider the length distribution in the case of a perfectly
transparent FS-interfaces, in which the length distribution is that of
one elementary block. Due to the roughness of the insulator and the
FS-interface the quasi-particles undergo diffusive reflection from these
boundaries. Incident and reflected directions are completely
uncorrelated. Then, assuming an uniform distribution for the directions
of ${\bf{v}_{ {F}}}$, we obtain for the length distribution of one
elementary block (corresponding to the case of $T=1$)
\begin{eqnarray}
   p_0(l)=\int_{0}^{1}d(\cos{\theta_i})\int_{0}^{1}d(\cos{\theta_r})
   \delta(l-\frac{d}{\cos{\theta_i}}-\frac{d}{\cos{\theta_r}}),
   \label{p0in}
\end{eqnarray}
where $\theta_i$($\theta_r$) denotes angle of the incident (reflected)
direction with respect to the normal to surface of the insulator. To
take into account the weak bulk disorder, we include a factor
$\exp{-l/\ell_{\text{imp}}}$. This serves mainly to yield a finite
average length of the distribution (\ref{p0in}). In a purely ballistic
layer with $p_0(l)$ given by (\ref{p0in}), the average length would
logarithmically divergent. Taking this into account we obtain
\begin{eqnarray}
   p_0(l)= \frac{2d}{Cl^2}\left[\frac{l-2d}{l-d}+
   \frac{2d}{l}\ln\frac{l-d}{d}\right]e^{-l/l_{\text{imp}}}
   \theta(\frac{l}{d}-2),
   \label{p0}
\end{eqnarray}
where $C= {E}^2 _2(d/\ell _{\text {imp}})$ ($ {E}_2(z) =
\int_{1}^{\infty}dx\exp{(-zx)}/x^2$ is the exponential integral of order
2).

In the second step, we connect the elementary building blocks, if the FS
interface has a transparency $T<1$. In determining the
length distribution we assume that an particle either goes through
the interface or is fully reflected. Only the number of these
reflection depends on $T$. We do not take into account quantum
mechanical interference for a single reflection at the
FS-interface. Taking this into account will lead essentially the
same results as our approach. By an expansion in the reflectivity
$R=1-T$ for the distribution $p(l)$ we can write
\begin{eqnarray}
  p(l)=T\sum_{n=0}^{\infty}R^n\int dl_0..dl_np_0(l_0)..p_0(l_n)
  \delta [l-\sum_{i=0}^{n}l_i]\,,
  \label{plr}
\end{eqnarray}
where the $n$th term in the expansion is the contribution of the
trajectories on which quasi-particles after $n$ times reflections
from FS-interface leave the F-layer. It is easy to see from Eq.
(\ref{plr}), that $p(l)$ obeys the integral equation
\begin{eqnarray}
  p(l)=Tp_0(l) +R\int dl^{\prime}p_0(l^{\prime})p(l-l^{\prime})\,,
  \label{pli}
\end{eqnarray}
which is readily solved by a Fourier transformation:
\begin{eqnarray}
  p(l)=\int_{-\infty}^{\infty}
  \frac{dk}{2\pi} e^{ikl}P(k).
  \label{plf}
\end{eqnarray}
Replacing (\ref{plf}) in Eq. (\ref{pli}) we find
\begin{eqnarray}
  P(k)=\frac{TP_0(k)}{1-R P_0(k)}\,,
  \label{pkr}
\end{eqnarray}
where $P_0(k)= {E}_2^2(ikd+d/\ell_{\text{imp}})/C$ is the Fourier
transform of $p_0(l)$.

\begin{figure}[h]
  \begin{center}
    \includegraphics[width=11cm,clip=true]{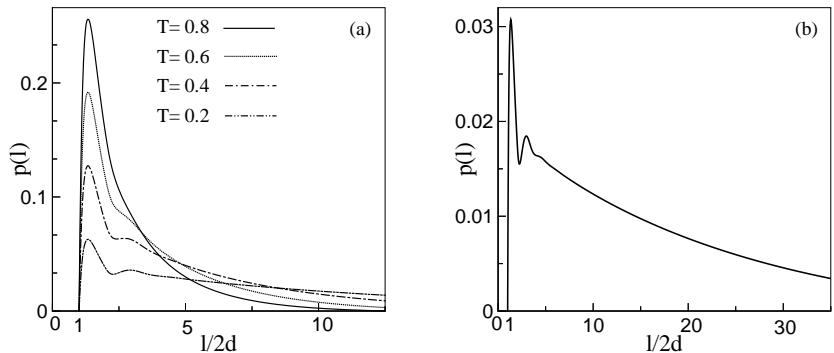}
    ~\\ ~\\
       \caption{The calculated distribution of the trajectory length in
      the F-layer: a) for different values of FS-transparency b) for
      small FS-transparency ($T=0.01$). The double peak structure close
      to the smallest length originates from the first two reflections,
      whereas the distribution for long trajectories decays as
      $\exp(-l/\bar{l})$, with the average length $\bar{l}\approx
      2d/T$.}
   \label{zbnfig2}
   \end{center}
\end{figure}

The distribution $p(l)$ determines the relevant length scale
associated with the geometrical size of the system corresponding
to the typical distances quasi-particles travel inside F. We have
plotted $p(l)$ for different $T$s and $d/\ell_{\text{imp}}=0.1$ in
Fig.~\ref{zbnfig2}. For small $T$, it has a characteristic double
peak structure close to the shortest trajectories $l\simeq 2d$,
resulting from trajectories reflected once and twice from the
insulator. At large $l$ the distribution decays exponentially as
$\exp(-l/\bar{l})$, where ${\bar
  l}\simeq 2d\ln(d/l_{\text{imp}})/T$ is the mean trajectory length. We
therefore have two characteristic lengths of the distribution, the
smallest possible trajectory length $2d$ and the average length
$\bar l$. The former determines the energy of the first Andreev
level and the latter the possible longest length of the
trajectories. For $T\sim 1$ these two length scale are of the same
order, leaving the thickness as the only relevant length scale. In
this case $p(l)$ has only one peak close to $2d$ (see
Fig.~\ref{zbnfig2}). Which of two length scales $2d$ and ${\bar
l}$ determines the total density of states will depend on the
other parameters.

Combining Eqs. (\ref{eq:dosb}) and (\ref{plf}) we obtain for the
total sub-gap DOS
\begin{eqnarray}
  N(E)=\frac{N_0}{2}\sum_{\sigma=\pm}
  \sum_{n=-\infty}^{\infty}P(k_n)
  e^{2ni\arccos{E/\Delta} },
  \label{dosb}
   \end{eqnarray}
where $k_n=2n(E+\sigma h)/v_{\text{F}}$. Similarly, from Eqs.
(\ref{eq:dosa}) and (\ref{plf}) the total DOS for the energies
above the gap is obtained:
\begin{eqnarray}
  N(E)=\frac{N_0}{2}\sum_{\sigma=\pm}
  \sum_{n=-\infty}^{\infty}P(k_n)
  e^{-2|n|{\text{acosh}} |E/\Delta |}.
  \label{dosa}
\end{eqnarray}
Thus, in both cases the density of states is fully expressed in
terms of the Fourier transform of the trajectory length
distribution. Most probably a real F-film  has a non-uniform
thickness due to the large scale roughness of the boundaries.
Assuming a smoothly varying thickness we can take this into
account by averaging expressions (\ref{dosb}) and (\ref{dosa})
over a Gaussian distribution of the thickness around a mean value
$d$. This will also leads to a smoothening of the sharp features
in DOS resulting from the lower cutoff in $p(l)$. The qualitative
behaviour will however not change. In our calculations, we have
taken a width of the distribution to be of order 10$\%$ that
correspond to condition of the experiments \cite{aprili:01}.

\section{Results and Discussions}
\label{sec:results}

Eqs.~(\ref{dosb}) and (\ref{dosa}) express the DOS of a F-layer
contacted by the superconductor in terms of the trajectory length
distribution. Depending on the relative values of $\Delta$, $h$,
the Thouless energy $v_{\text F}/2d$ and $T$, the resulting DOS
has different behaviors. We will concentrate mainly on the limit
$T\ll1$. For the length dependence it useful to distinguish
between normal metal with $h=0$, a weak ferromagnet $h\sim\Delta$
of $h<\Delta$ and $h>\Delta$ and a strong ferromagnetic film of
$h\gg\Delta$ . We analyze the DOS in each case for different
values of $d_T \Delta /v_{\text{F}}$ ($2d_T={\bar l}$), being the
relevant length scale in the limit $T\ll1$.  In the end, we
summarize all results in a map of the DOS depending on $d_T \Delta
/v_{\text{F}}$ and $d_T h /v_{\text{F}}$.

\subsection{Normal film}
\label{subsec:normal}

\begin{figure}
  \begin{center}
    \includegraphics[width=17cm,clip=true]{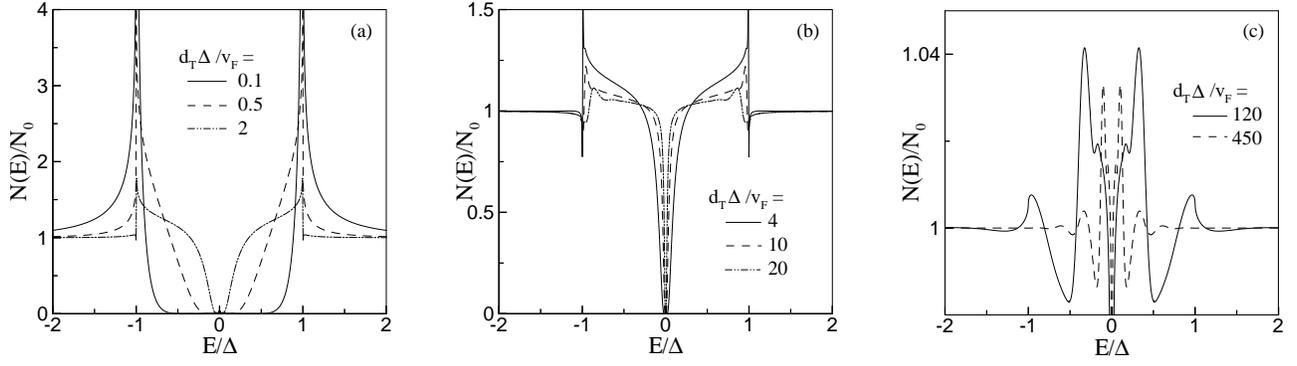}
    ~\\ ~\\
    \caption[]{The DOS vs. energy of the normal layer ($h=0$) for
      different values of $d_T\Delta/v_{\text {F}}$. a) Suppression of the
      superconducting features of DOS with increasing $d_T\Delta/v_{\text
        {F}}$ and appearance of a minigap of order $v_{\text {F}}/d_T$.  b)
      The first Andreev peaks (AP) at roughly $\pm v_{\text F}/2d$ correspond
      to the first peak in the distribution of the trajectory length.  c) For
      very large $d_T\Delta/v_{\text {F}}$ the DOS has many AP, leading to
      small deviations of order $T$ from the normal state DOS. }
   \label{zbnfig3}
   \end{center}
\end{figure}

Let us start with a normal metal film ($h=0$) contacted by the
superconductor. The DOS is shown in Fig.~\ref{zbnfig3}a-c, for
different values of $d_T \Delta /v_{\text{F}}$. In the limit of a
very thin layer with $d_T \Delta /v_{\text{F}}\ll 1$, the DOS has
essentially the form of superconducting DOS with sharp peaks at
$E=\pm \Delta$ and zero DOS for energies inside the gap. By
increasing $d_T \Delta /v_{\text{F}}$ the peaks are getting
broader and a finite DOS appears at small energies.  There is
still an energy interval around $E=0$ with zero DOS (see
Fig.~\ref{zbnfig3}a).  Increasing $d_T \Delta/T$ further leads to
a suppression of the superconducting features of the DOS. The zero
DOS interval become smaller and the DOS at other energies tends to
be closer to the DOS of the normal state. Thus, the density of
states develops a minigap around the Fermi level, which decreases
with increasing $d_T$ roughly as $v_{\text{F}}/d_T =
v_{\text{F}}T/d\ln{(\ell_{\text {imp}}/d)}$ (see Fig.
~\ref{zbnfig3}b).  This minigap is related to the mean length of
the trajectories ${\bar l}$, which has a finite value, if
$d/\ell_{\text
  {imp}}$ is finite. The presence of weak bulk disorder in the normal
film suppresses long trajectories. Formally, this was included in the
distribution of the trajectory length as the exponentially decaying
factor in Eq. (\ref{p0}), which leads to the finite mean length ${\bar
  l}=2d\lg{\ell_{\text {imp}}/d}$. This act as an effective upper limit
of order ${\bar l}$ of the length of the trajectories, which gives
a lower bound to the energy of the Andreev bound states. Similar
features were found before within a tunneling model
\cite{mcmillan:68} and in the diffusive models
\cite{golubov,Belzigthesis} in a disordered normal layer contacted
by the superconductor.

The peaks of $E\pm=\Delta$ originate from the first peak in the
distribution $p(l)$ at $l\simeq 2d$, which at higher $d_T \Delta
/v_{\text{F}}$ move to lower energies given roughly by $\pm
v_{\text{F}}/2d$ (see Fig.~\ref{zbnfig3}b). They originate from
Andreev peaks (AP) resulting from the trajectories with $l\simeq
2d$. By increasing $d_T \Delta /v_{\text{F}}$ the first AP moves
to lower energies, and, when $d_T \Delta /v_{\text{F}}\gtrsim
1/T$, the next AP appear at $E\pm=\Delta$ (Fig. \ref{zbnfig3}c).
In this case the DOS is close to the normal states values. Small
deviations proportional to $T$ display many AP, as shown in
Fig.~\ref{zbnfig3}c. Small peaks close the main peaks, which are
more pronounced for the first AP, correspond to the second peak of
$p(l)$ at $l\simeq 4d$.

\subsection{Very weak ferromagnet, $h<\Delta$}
\label{subsec:weak}

Now we study the effect of the spin splitting in a ferromagnetic
film on the DOS. First we consider the case of a weak ferromagnet,
where the exchange field is of order of the superconducting gap,
but $h<\Delta$. A qualitative picture of the influence of $h$ on
the DOS follows from the condition (\ref{quantcond}) for the
formation of Andreev bound states in the ferromagnetic layer. This
is the semiclassical quantization condition for coherent
superposition of two subsequent Andreev reflection of a
quasi-particle from S, which propagate along a trajectory of
length $l$. It includes the phase gained by a quasiparticle of
spin $\sigma$ along a trajectory $(E+\sigma h)l$, and the phase
shift produced by an Andreev reflection, $\arccos{(E/\Delta)}$. If
the energy of the quasi-particle is not close to $\Delta$ the
phase shift resulting from Andreev reflection is different from
zero. To obtain constructive interference the total phase must be
an integer multiple of $2\pi$, as follows from
Eq.~(\ref{quantcond}).  The existence of an upper limit on the
length of the trajectories (as discussed above), leads to the
formation of a zero DOS interval (minigap) around $E=\pm h$.
Therefore, the total subgap DOS should be similar to the average
of two by $\pm h$ shifted normal spectra.

In column (I) of Fig.~\ref{zbnfig4} the DOS of the ferromagnetic
film with $h=0.5\Delta$ is shown for different values of $d_T
\Delta /v_{\text{F}}$. Shifting the minigap leads to minima at
$E=\pm h$. The zero energy DOS becomes finite and increases with
increasing $d_T \Delta /v_{\text{F}}$. At higher $d_T \Delta
/v_{\text{F}}\sim 1$ the DOS has a smooth peak at zero energy and
two dips at $E\pm h$ (see Fig. \ref{zbnfig4}-Ia). Here, the AP are
located at $E\pm \Delta$. By increasing $d_T \Delta /v_{\text{F}}$
the width of the dips decreases roughly as $v_{\text{F}}/d_T$ and
the first AP moves to energies below the gap, i.~e., $\sim\pm
v_{\text{F}}/d$ (see Fig.~\ref{zbnfig4}-Ib). We can distinguish
two domains of energies below the exchange field $|E|<h$ and
energies above the exchange field $|E|>h$. At higher $d_T \Delta
/v_{\text{F}}$ the first AP moves from $|E|>h$ to $|E|<h$ and the
next AP appears at $|E|>h$. For the region $|E|<h$ the DOS shows a
zero energy peak, if the first APs merge at $E=0$ (see Fig.
\ref{zbnfig4}-Ic). This is a zero energy Andreev peak (ZEAP),
which originates from phase shifting caused by the exchange field.
Additional shifting of the AP results in an oscillatory behavior
of the DOS in the domain $|E|<h$. As shown in
Fig.~\ref{zbnfig4}-Id, in the limiting case of $d_T \Delta
/v_{\text{F}}\gg 1/T$ the DOS is close to the normal state value,
exhibiting small deviations, which are of the order of the
FS-interface transparency $T$. The deviations have the form of
small oscillations at all energies.

\subsection{Weak ferromagnet, $h>\Delta$}
\label{subsec:weak2}

\begin{figure}
  \begin{center}
    \includegraphics[width=17cm,clip=true]{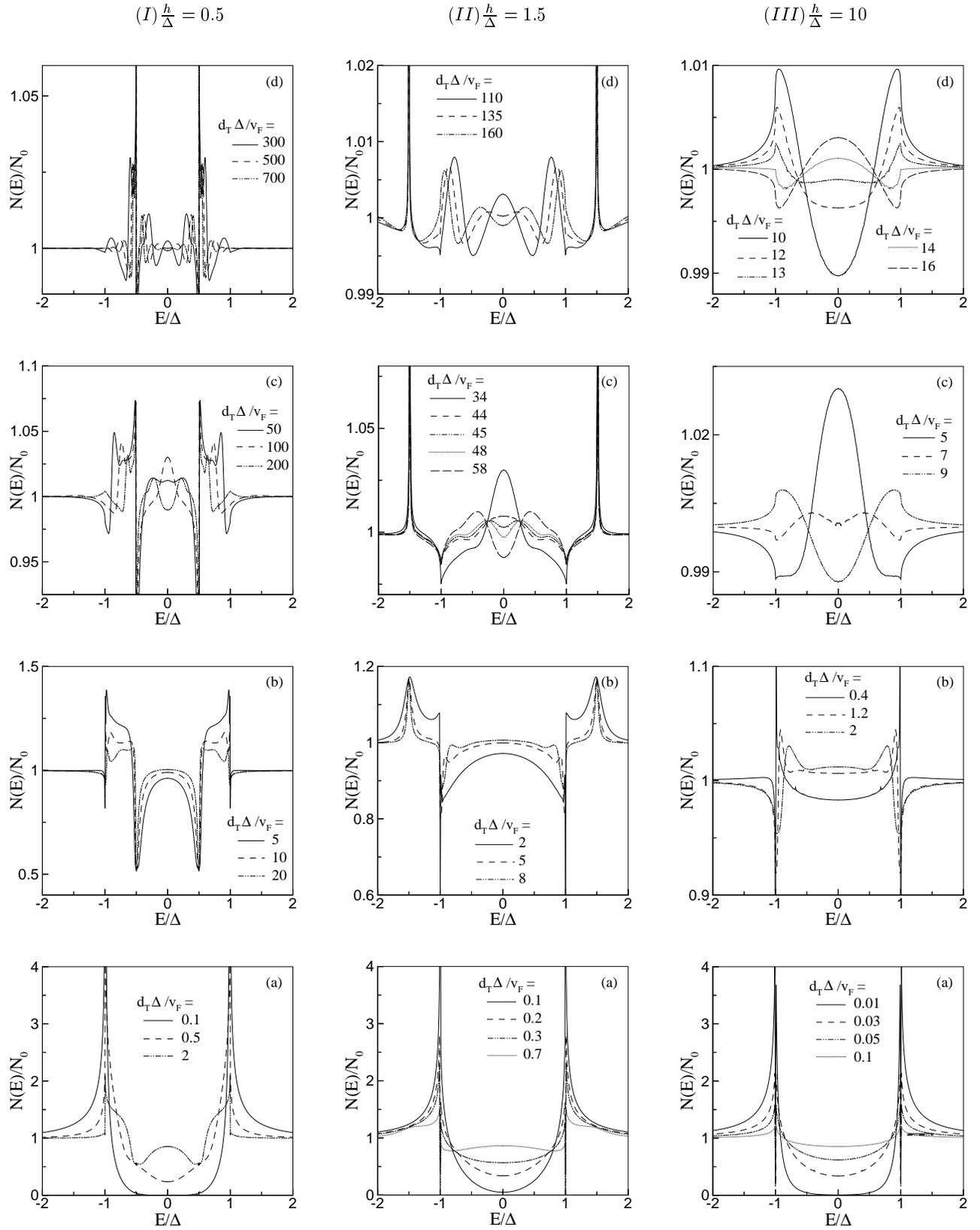}
    \caption[]{
      Density of states in a ferromagnetic layer in contact to a
      superconductor. Different columns labeled (I)-(III) correspond to
      the values of the exchange field given above. The thickness of the
      layer is increased from the bottom to the top. For explanation of
      the various regimes, see the text. }
       \label{zbnfig4}
       \end{center}
\end{figure}

In the case of $h>\Delta$ the suppression of the superconducting
features of DOS occurs at lower $d_T \Delta /v_{\text{F}}$,
compared with the previous case of $h<\Delta$ (see
Fig.~\ref{zbnfig4}-IIa). The subgap DOS has similar features as it
had for energies below $h$ in the case of $h<\Delta$. The zero
energy DOS increases to the normal state values as
$d_Th/v_{\text{F}}$ becomes of order unity.  Then, the DOS has a
smooth peak at $E=0$ and minima at $E\pm \Delta$, as is shown in
Fig.~\ref{zbnfig4}-IIb-c. The AP at $E\pm \Delta$ move to lower
energies at a higher values of $d_T \Delta /v_{\text{F}}$ and form
ZEAP, when they merge at $E=0$ (see Fig.~\ref{zbnfig4}-IIc ). The
size of the ZEAP is of order of $T$.

The main feature of DOS in energies above the gap consists of
sharp peaks at $E=\pm h$ (see Fig.~\ref{zbnfig4}-IIb-d ). These
peaks originate from a resonant transmission of the
quasi-particles through the superconducting potential ($\Delta=0$)
inside the F-film. A quasi-particle incident from the
superconductor to the F-film with energy above the gap, is

scattered by the potential well whose width is determined by the
length of the quasi-particle trajectory $l$ inside F. The phase
gained by the quasi-particle of spin $\sigma$ inside the potential
well is given by $(E+\sigma h)l$. At $E=-\sigma h$ the incident
and transmitted quasi-particles interfere constructively, which
leads to a reflection-less transmission. Similar effect were found
before in normal metal-superconductor hetrostructures
\cite{reflectionless}. For $d_Th/v_{\text{F}}>1/T$ the subgap DOS
shows an oscillatory behavior around the normal state value as a
function of $dh/v_{\text{F}}$ (see Figs.\ref{zbnfig4}-IIc and
\ref{zbnfig4}-IId). The period of the oscillation is $\pi/2$ and
the amplitude is of order $T$. The amplitude is damped in the
limit of $d_T\Delta/v_{\text{F}}\gg1/T$. Note that the phase of
the oscillation depends on the energy.

\subsection{Ferromagnetic film}
\label{subsec:strong}

Now let us consider rather strong exchange fields $h\gg\Delta$. In
column III of Fig.~\ref{zbnfig4} we plotted $N(E)$ for $h=10
\Delta$ at different $d_T \Delta /v_{\text{F}}$. As shown in
Fig.~\ref{zbnfig4}-IIIa the suppression of the superconducting
features from DOS by increasing $d_T \Delta /v_{\text{F}}$ is
faster than in the weak ferromagnetic cases. In fact the DOS
reaches almost to the normal state value for $d_T h
/v_{\text{F}}\sim 1$.  As long as $d_T h /v_{\text{F}}\lessim 1$,
mainly long trajectories of $l\sim {\bar l}$ contribute to the
zero energy DOS, which are well approximated by an exponential
distribution of the form $\exp{(-l/{\bar l})}$.  Replacing this
approximation form of $p(l)$ into the general expression of DOS
Eq.~(\ref{eq:dosb}) we find, that the zero energy DOS increases
with $d_T h /v_{\text{F}}$ roughly as
\begin{eqnarray}
   N(0)=N_0\frac{\pi v_{\text F}}{h{\bar l}}
   \frac{\exp{(-\pi v_{\text F}/2h{\bar l})}}
   {1-\exp{(-\pi v_{\text F}/h{\bar l})}}\;.
\label{0dos}
\end{eqnarray}
This results is also applicable for the weak ferromagnetic case
discussed above. As before, the shifting of the AP to lower
energies (see Fig.~\ref{zbnfig4}-IIIb) leads to the formation of
ZEAP at the Fermi level as is seen from Fig.~\ref{zbnfig4}-IIIc/d.
Then the DOS develops coherent oscillation as a function of
$dh/v_{\text{F}}$ with the period $\pi/2$. The amplitude and the
sign of the oscillation depend on energy.  Maximal amplitudes of
opposite sign always occur at zero energy or at the gap energy
(see Fig.~ \ref{zbnfig4}-IIId). This results in an inverted energy
dependence of the DOS by changing $d$, which has been observed in
the experiment \cite{aprili:01}. We have shown in \cite{ZBN:01}
that our results is in a quantitative agreement with the
experimental data.

\subsection{Maps of the proximity DOS}
\label{subsec:map}

Summarizing the above analysis we present a map showing the
dependence of DOS on $d_T \Delta /v_{\text{F}}$ and $d_T h
/v_{\text{F}}$ for small FS-interface transparency. This map is
shown in Fig.~\ref{zbnfig5}. Various regions in the map are
distinguished by different ranges of $h/\Delta$ and $d_T \Delta
/v_{\text{F}}$ (or equivalently $d_T h /v_{\text{F}}$). Along the
diagonal lines $h$ is equal to $\Delta$ and moving upwards $d_T
\Delta /v_{\text{F}}$ (and consequently $d_T h /v_{\text{F}}$)
increases.  The quarter circles are curves with constant $d_T$,
along which the ratio $h/\Delta$ is varying.  In the following we
discuss different regions according to this classification. The
normal film corresponds to the vertical axis ($h=0$), which
consists of three parts. The first part is limited by $d_T \Delta
/v_{\text{F}}\lessim 1$. Here the superconducting features are
dominant at lower $d_T \Delta /v_{\text{F}}$ and suppressed for
$d_T \Delta /v_{\text{F}}\sim 1$, showing a minigap at the Fermi
level. The second part is limited by $1\grtsim d_T \Delta
/v_{\text{F}}\lessim 1/T$, where the main feature is the first AP
at energies $\sim\pm v_{\text{F}}/2d$ and a minigap of order
$v_{\text{F}}/d_T$. Close to the boundary $ d_T \Delta
/v_{\text{F}}\sim 1/T$ the second AP appears in the DOS. Finally,
the third part is the region $d_T \Delta /v_{\text{F}}>1/T$, where
the DOS contains many AP, appearing as small deviations
(proportional to $T$) from the normal state DOS.

In the case of a ferromagnetic film we distinguish the following
regions:

\begin{enumerate}

\item[i)] The strongly superconducting region is limited by the smallest
  quarter circle, in which the superconducting features are dominate the
  DOS. At non-zero $h$ the zero energy DOS appears at larger $d_T
  \Delta/v_{\text{F}}$ and the DOS increases to the value of the normal
  state in the domain close to the second quarter circle boundary. For
  the part $h<\Delta$, there is a smooth maximum at $E=0$ between two
  minima at $E=\pm h$.  For $h>\Delta$ we have only a smooth peak at
  $E=0$.

\item[ii)] The intermediate regions limited by two quarter circles. In
  $h<\Delta$ part the main features in DOS is the existence of two dips
  at $E\pm h$ and the first AP. Close to the second boundary we observe
  a separation between two energy domains $|E|<h$ and $|E|>h$. While the
  second AP peaks appears at energies above $h$, the first AP peaks move
  to energies below $h$ domain. In the $h>\Delta$ part the DOS has a
  smooth peak at $E=0$. In the domain close to the diagonal line
  $h \grtsim \Delta$, there are also two resonance sharp peaks at $E\pm h$
  which disappear in $h\gg \Delta$ regions. In both cases of $h<\Delta$ and
  $h>\Delta$, shifting of the APs leads to the formation of a ZEAP.
  This happens at regions close to the third quarter circle.

\begin{figure}
  \begin{center}
    \includegraphics[width=14cm,clip=true]{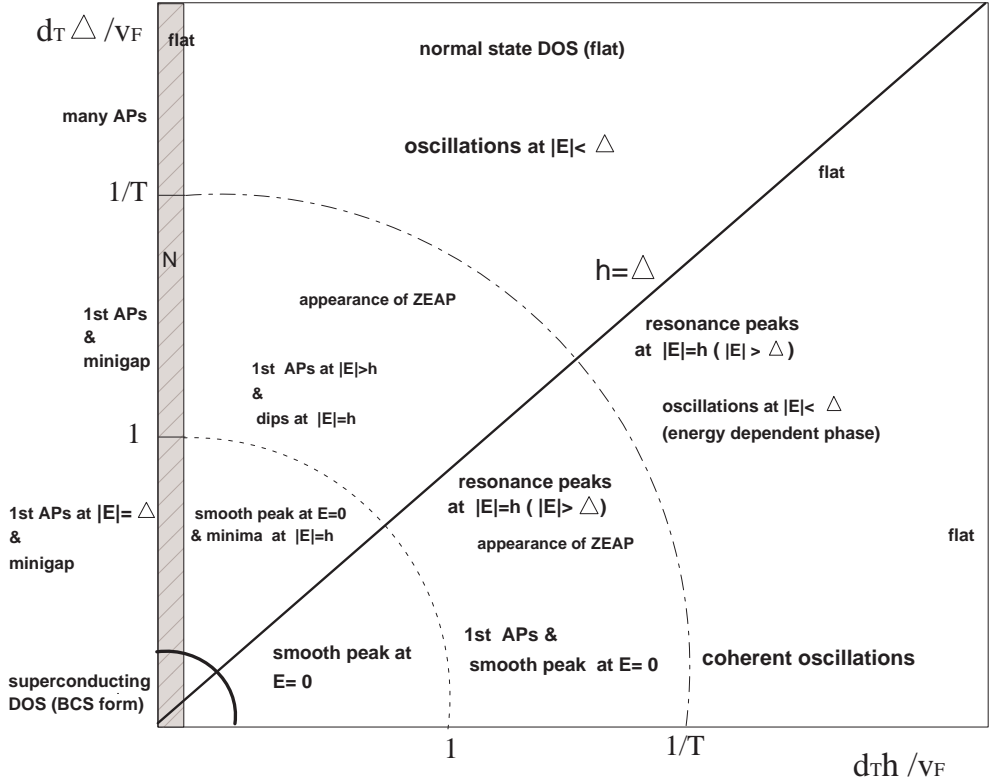}
    \caption[]{
      Map of the proximity DOS in a ferromagnetic layer showing the
      dependence on $d_T \Delta/v_{\text{F}}$ and $d_T h/v_{\text{F}}$ for a
      small FS-interface transparency $T\ll 1$. In the case of high
      transparency $T\sim 1$ the region between $1$ and $1/T$ is absent.}
      \label{zbnfig5}
  \end{center}
\end{figure}

\item[iii)] In the region above the third quarter circle the DOS is
  close to the normal state value (flat).  There are, however, small
  deviations proportional to $T$, which have different origins in the
  different domains. In the limiting domains of $h\ll\Delta$ and
  $h\gg\Delta$ (close to the respective axis') they consist of many AP
  and the oscillations, respectively. In the $h<\Delta$ part there are
  many AP above $h$ and an oscillatory variation below $h$. In the
  $h>\Delta$ part and
  for weak exchange fields in subgap part we have oscillatory variations
  with an energy dependent sign and amplitude, which result from a
  collective shift of many AP. As the intermediate region the resonance
  peaks are present at $E=\pm h$. In the strong exchange field part
  these peaks are disappear. Here the oscillations of subgap DOS
  are produced by the first AP only.
\end{enumerate}

The DOS map for the case of high transparency $T\sim 1$ is similar
to Fig.~\ref{zbnfig5}. The difference is the absence of the region
between $1$ and $1/T$ on both axis.  In the remaining regions we
have features similar to the $T\ll1$ case. The value of the
minigap and the energy of first AP are the same order. All
features and variations including AP and oscillations are more
pronounced than the $T\ll1$ case.

\section{Conclusion}
\label{sec:conclusion}

In summary, we have studied theoretically superconducting
proximity effect in ballistic ferromagnetic layers. Within the
quasi-classical formalism, we have obtained expressions for the
density of states at all energies in a ferromagnetic metal in
contact with a superconductor which are completely specified by
Fourier transform of the length distribution of classical
trajectories in the ferromagnet. The length distribution of
trajectories depends only on the geometrical properties of the
attached ferromagnet and the connected interfaces. Thus, the
obtained expressions are applicable for ballistic FS structures of
arbitrary geometry. We have calculated the length distribution for
the film geometry in a quasi-ballistic model taking into account
finite transparency of the FS interface, roughness at the film
boundaries and weak bulk disorders. The density of states exhibits
variety of structures depending on the values of the
superconducting energy gap $\Delta$, the ferromagnet exchange
field $h$, the thickness of F film $d$ and the FS interface
transparency $T$.  We have observed many interesting features,
like splitting of the subgap density of states for spin up and
spin down quasi-particles, a zero energy Andreev peak, resonant
transmission peaks above the gap at $E=\pm h$ and oscillations of
the DOS as a function of the film thickness. These effects have
been explained in terms of the phase shift of Andreev bound
states, caused by the exchange field. We have analyzed the density
of states in the full parameter range and summarized the results
in a map, shown in Fig.~\ref{zbnfig5}.

\acknowledgements One of the authors (M. Z.) would like to thank
the group of Prof. Bauer at Delft University of Technology for
their hospitality and support during his visit.


\begin{thebibliography}{99}
\bibitem{rowell}
  W.~L.~McMillan and J.~R.~Rowell, in {\em Superconductivity}, edited by
  R.~D. Parks (1969, Marcel Dekker, New York).

\bibitem{mcmillan:68}
  W.~L.~McMillan, Phys. Rev. {\bf 175}, 537 (1968).

\bibitem{andreev}
  A.~F. Andreev,
  Sov. Phys. JETP {\bf 19}, 1228 (1964).

\bibitem{gueron}
  S. Gu\'eron, H. Pothier, N.~O. Birge, D. Esteve, and M. H. Devoret,
  Phys. Rev. Lett. {\bf 77}, 3025 (1996).

\bibitem{belzig:96}
  W. Belzig, C. Bruder, and G. Sch\"on,
  Phys. Rev. B {\bf 54}, 9443 (1996).

\bibitem{tessmer}
  S.~H. Tessmer, M.~B. Tarlie, D.~J. Van Harlingen, D.~L. Maslov, and
  P.~M. Goldbart, Phys. Rev. Lett. {\bf 77}, 924 (1996).

\bibitem{millo}
  Y. Levi, O. Millo, N. D. Rizzo, D.~E. Prober, and L.~R.
  Motowidlo, Phys. Rev. B {\bf 58}, 15128 (1999).

\bibitem{courtois}
  N. Moussy, H. Courtois, and B. Pannetier, Rev. Sci.
  Instrum. {\bf 72}, 128 (2001).

\bibitem{vinet}
  M. Vinet, C. Chapelier, and F. Lefloch,
  Phys. Rev. B {\bf 63}, 165420 (2001).


\bibitem{falko}
  V.~I.~Fal'ko, C.~J.~Lambert, A.~F.~Volkov,
  JETP Lett.~{\bf 69}, 532-538 (1999);

\bibitem{jedema}
  F.~J.~Jedema, B.~J.~van Wees, B.~H.~Hoving, A.~T.~Filip, and
  T.~M.~Klapwijk, Phys.~ Rev. B {\bf 60}, 16549 (1999).

\bibitem{belzig:00}
  W. Belzig, A. Brataas, Yu.~V. Nazarov, and G.~E.~W. Bauer,
  Phys. Rev. B {\bf 62}, 9726 (2000).

\bibitem{tedrow}
  W.~J.~Gallagher, D. E. Paraskevopoulos, P. M. Tedrow,
  S. Frota-Pessoa, and B. B. Schwartz, Phys. Rev. B {\bf 21},
  962 (1980); see also R. Meservey
  and P.~M.~Tedrow, Phys. Rep. {\bf 238}, 173 (1994).

\bibitem{aprili:01}
  T. Kontos, M. Aprili, J. Lesueur and X. Grison,
  Phys. Rev. Lett. {\bf 86}, 304(2001).

\bibitem{blatter}
  L. B. Ioffe, V. B. Geshkenbein, M. V. Feigel'man, A. L. Fauchere,
  and G. Blatter,  Nature {\bf 398}, 679 (1999).

\bibitem{mooij}
  J.~E. Mooij, T.~P. Orlando, L. Levitov, L. Tain, C.~H. van der
  Wal, and S. Lioyd,
  Science {\bf 285}, 1036  (1999).

\bibitem{scheer:97}
  E. Scheer, P. Joyez, D. Esteve, C. Urbina, and M.~H. Devoret,
  Phys. Rev. Lett. {\bf 78}, 3535 (1997).

\bibitem{scheer:98}
  E. Scheer, N. Agrait, J.~C. Cuevas, A. Levy Yeyati, B. Ludoph,
  A. Martin-Rodero, G.~R. Bollinger, J.~M. Van Ruitenbeek, and C. Urbina,
  Nature {\bf 394}, 154 (1998).

\bibitem{scheer:01}
  E. Scheer, W. Belzig, Y. Naveh, C. Urbina, D. Esteve, and M. H. Devoret,
  Phys. Rev. Lett. {\bf 86}, 296 (2001).

\bibitem{LO}
  A. I. Larkin, and Yu. N. Ovchinnikov, Sov. Phys.
  JETP {\bf 20}, 762 (1995).

\bibitem{FF}
  P. Fulde, and R.~A. Ferrel, Phys. Rev. {\bf 135}, A550 (1964).

\bibitem{Radovic}
  Z. Radovic, M. Ledvij, L. Dobrosavljevic-Grujic, A. I. Buzdin,
  and J.R. Clem , Phys. Rev. B {\bf 44}, 759 (1991).

\bibitem{strunk}
  J.~S. Jiang , D. Davidovic, Daniel H. Reich, and C. L. Chien
  , Phys. Rev. Lett. {\bf 74}, 314 (1995);
  Th. M\"uhge, N. N. Garif'yanov, Yu. V. Goryunov,
  G. G. Khaliullin, L. R. Tagirov, K. Westerholt,
  I. A. Garifullin, and H. Zabel,
  {\it ibid}. {\bf 77}, 1857 (1996).

\bibitem{aarts} J. Aarts {\it et al.},
 Phys. Rev. B {\bf 56}, 2779 (1997);
 L. Lazar {\it et al.},
 {\it ibid.}, {\bf 61}, 3711 (2000);
 M. Sch\"ock, C. S\"urgers,
 and H. v. L\"ohneysen, Eur. Phys. J. B {\bf 14}, 1 (2000).

\bibitem{ryasanov}
  V.~V.~Ryazanov, A.~V. Veretennikov, V. A. Oboznov, A. Yu. Rusanov,
  V. A. Larkin, A. A. Golubov, and J. Aarts,
  Physica B {\bf 284-288},495 (2000);
  V.~V.~Ryazanov, V. A. Oboznov, A. Yu. Rusanov, A.~V. Veretennikov,
  A. A. Golubov, and J. Aarts,
  Phys. Rev. Lett. {\bf 86}, 2427 (2001).

\bibitem{bulaevski}
   L. N. Bulaevskii, V. V. Kuzii, and A. A. Sobyanin, Pis'ma Zh.
   Eksp. Teor. Fiz. {\bf 25}, 314 (1977)[JETP Lett. {\bf 25}, 290
   (1997)].

\bibitem{ZBN:01}
   M. Zareyan, W. Belzig, and Yu. V. Nazarov,
   Phys. Rev. Lett. {\bf 86}, 308 (2001).

\bibitem{pilgram:00}
   S. Pilgram, W. Belzig, and C. Bruder, Phys. Rev. B
   {\bf 62}, 12462 (2000).

\bibitem{eilenberger}
  G. Eilenberger,
  Z. Phys. {\bf 214}, 195 (1968);
  A.~I. Larkin and Yu.~N. Ovchinnikov,
  Sov. Phys. JETP {\bf 26}, 1200 (1968).

\bibitem{KO}
   I. O. Kulik, A. N. Omelyanchouk, Sov. J. Low Temp. Phys. {\bf 4},
   142 (1978).

\bibitem{golubov}
  A.~A. Golubov and M.~Yu. Kupriyanov,
  J. Low Temp. Phys. {\bf 70}, 83 (1988);
  A.~A. Golubov and M.~Yu. Kupriyanov,
  Sov. Phys. JETP {\bf 69}, 805 (1989).

\bibitem{Belzigthesis}
   W. Belzig , PhD thesis, Karlsruhe university (1999).

\bibitem{reflectionless} A. Kastalsky, A. W. Kleinsasser,
L. H. Greene, R. Bhat, and J. P. Harbison,
 Phys. Rev. Lett. {\bf 67}, 3026 (1991);
 C. Nguyen, H. Kroemer, and E. L. Hu,
 Phys. Rev. Lett. {\bf 69}, 2847 (1992);
 see also M. Schechter, Y. Imry, Y. Levinson,
 cond-mat/0104343(unpublished).

\end{thebibliography}
\end{document}